\documentclass[aps,prx,groupedaddress,showpacs,twocolumn,nofootinbib,superscriptaddress,notitlepage, longbibliography]{revtex4-1}
\usepackage{times}
\usepackage{amsmath,empheq}
\usepackage{amssymb,bm}
\usepackage{amsthm,color,xcolor,dsfont}
\usepackage{graphicx} 
\usepackage{xcolor}
\usepackage{epstopdf}
\usepackage[colorlinks=true, hyperindex, breaklinks, linkcolor=blue, urlcolor=blue, citecolor=blue]{hyperref} % Physical Review style
\usepackage{ulem}
\newcommand{\CC}{\mathcal{C}} 
\newcommand{\DC}{\mathcal{D}}
\newcommand{\EC}{\mathcal{E}}

\newcommand{\OC}{\mathcal{O}}

\newcommand{\RC}{\mathcal{R}}

\newcommand{\ket}[1]{|#1\rangle} % ket
\long\def\ca#1\cb{} %Use for commenting out: \ca...\cb

\newcommand\add[1]{{#1}} % remove the \add marking

\newcommand\delete[1]{\unskip} % delete

\newtheorem*{definition*}{Definition}

\begin{document}

\title{On the robustness of bucket brigade quantum RAM}

\author{Srinivasan Arunachalam} 
\email{S.Arunachalam@cwi.nl} 
\affiliation{Centrum Wiskunde \& Informatica (CWI), Amsterdam, The Netherlands}
\affiliation{Institute for Quantum Computing, University of Waterloo, Waterloo, ON, N2L 3G1, Canada}

\author{Vlad Gheorghiu} 
\email{vlad.gheorghiu@uwaterloo.ca}
\affiliation{Institute for Quantum Computing, University of Waterloo, Waterloo, ON, N2L 3G1, Canada}
\affiliation{Department of Combinatorics \& Optimization, University of Waterloo, Waterloo, ON, N2L 3G1, Canada}

\author{Tomas Jochym-O'Connor} 
\email{trjochymoconnor@uwaterloo.ca} 
\affiliation{Institute for Quantum Computing, University of Waterloo, Waterloo, ON, N2L 3G1, Canada}
\affiliation{Department of Physics \& Astronomy, University of Waterloo, Waterloo, ON, N2L 3G1, Canada}

\author{Michele Mosca} 
\email{michele.mosca@uwaterloo.ca}
\affiliation{Institute for Quantum Computing, University of Waterloo, Waterloo, ON, N2L 3G1, Canada}
\affiliation{Department of Combinatorics \& Optimization, University of Waterloo, Waterloo, ON, N2L 3G1, Canada}
\affiliation{Perimeter Institute for Theoretical Physics, Waterloo, ON, N2L 6B9, Canada}
\affiliation{Canadian Institute for Advanced Research, Toronto, ON,  M5G 1Z8, Canada}

\author{Priyaa Varshinee Srinivasan} 
\email{priyaavarshinee.srin@ucalgary.ca} 
\affiliation{David R. Cheriton School of Computer Science, University of Waterloo, Waterloo, ON, N2L 3G1, Canada}
\affiliation{Department of Computer Science, University of Calgary, Calgary, AB, T2N 1N4, Canada}
\affiliation{Institute for Quantum Science and Technology, University of Calgary, Calgary, AB, T2N 1N4, Canada}

%\date{Version of \today}

\begin{abstract} 
We study the robustness of the bucket brigade quantum random access memory model introduced by Giovannetti, Lloyd, and Maccone~\href{http://dx.doi.org/10.1103/PhysRevLett.100.160501}{[Phys. Rev. Lett. \textbf{100}, 160501 (2008)]}. Due to a result of  Regev and Schiff~\href{http://link.springer.com/chapter/10.1007\%2F978-3-540-70575-8_63}{[ICALP '08 pp. 773]}, we show that for a class of error models the error rate per gate in the bucket brigade quantum memory has to be of order $o(2^{-n/2})$ (where $N=2^n$ is the size of the memory) whenever the memory is used as an oracle for the quantum searching problem. We conjecture that this is the case for any realistic error model that will be encountered in practice, and that for algorithms with super-polynomially many oracle queries the error rate must be super-polynomially small, which further motivates the need for quantum error correction. \add{By contrast, for algorithms such as matrix inversion~\href{http://dx.doi.org/10.1103/PhysRevLett.103.150502}{[Phys. Rev. Lett. \textbf{103}, 150502 (2009)]} or quantum machine learning~\href{http://dx.doi.org/10.1103/PhysRevLett.103.150502}{[Phys. Rev. Lett. \textbf{113}, 130503 (2014)]} that only require a polynomial number of queries, the error rate only needs to be polynomially small and quantum error correction may not be required.}
We introduce a circuit model for the quantum  bucket brigade architecture and argue that quantum error correction for the circuit causes the quantum bucket brigade architecture to lose its primary advantage of a small number of ``active" gates, since all components have to be actively error corrected. 
\end{abstract}

%\pacs{03.67.Pp, 03.67.Lx, 03.67.-a}
\maketitle

%%%% Introduction %%%%

\section{Introduction\label{sct:Introduction}}
A random access memory (RAM) is a 
device that stores information in an array of memory cells in the form of bits. In contrast to other
types of information storage devices, the access latency to any memory cell is
constant and does not depend on the location of the information in the RAM.
Information stored in the RAM is retrieved by inputting the address of the desired memory 
cell in a routing circuit. 
Any address in a RAM with $N=2^n$ memory cells can by addressed via a unique $n$ bit input query string.
The corresponding output  register contains the contents of the addressed memory location. 

The typical physical implementation of the addressing mechanism uses the fanout architecture
\cite{MicroCircDesign,MicroCirc}, in which the routing scheme corresponds to a
binary tree. Each node consists of a pair of transistors which routes the electronic signal down one of the two paths to the subsequent level. In the fanout architecture, a given level has all nodes sharing the same routing direction (left or right), set by the corresponding address bit. An $n$~bit query string determines a unique path in the binary tree, corresponding to the desired memory location. In the process, $\OC(2^n)$ transistors are activated.
 
Alternative routing schemes  with $\OC(\text{poly}(n))$ activated transistors have been proposed, corresponding to exponentially lower energy consumption. One such example is the ``bucket brigade" scheme \cite{PhysRevLett.100.160501, PhysRevA.78.052310}.
However, most of the classical implementations follow the simpler fanout architecture, as the power consumption of RAM is negligible in comparison with the power consumption of other components in the architecture.

The classical RAM addressing scheme can be generalized to a quantum RAM (which we simply call qRAM from here on) scheme, where the input is a quantum state, the
routing components are inherently quantum, and the information  stored can be either classical, i.e. $\ket{0}$ or $\ket{1}$ but not a superposition of both, or quantum, i.e. any arbitrary superposition of $\ket{0}$ and $\ket{1}$. 
In the present paper we consider qRAM that stores only classical information. 
Such memory allows querying superposition of addresses
\begin{equation}\label{eqn:1} 
\sum_j\alpha_j\ket{j}\ket{0} \stackrel{qRAM}{\longrightarrow} \sum_j\alpha_j\ket{j}\ket{m_j},
\end{equation}
where $\sum_j\alpha_j\ket{j}$ is a superposition of queried addresses and $\ket{m_j}$ represents the content of the $j$-th memory location. A memory that stores classical information but allows queries in superposition is required for quantum algorithms such as Grover's search on a classical database \cite{quantph.9605043}, collision finding \cite{Brassard:1997:QCH:261342.261346}, element distinctness \cite{Ambainis:2007:QWA:1328722.1328730}, dihedral hidden subgroup problem \cite{Kuperberg:2005:SQA:1085579.1093661}, \add{quantum matrix inversion~\cite{PhysRevLett.103.150502, PhysRevLett.109.050505}, quantum machine learning~\cite{quantph.1307.0411, PhysRevLett.113.130503, Nature.10.631, quantph.1408.3106} } and various practical applications mentioned in \cite{arunachalam2014quantum}. In fact, such a quantum memory plays the role of the oracle and is ideal in implementing any oracle-based quantum algorithm, in which the oracle is used to query classical data in superposition. \add{It is important to distinguish between algorithms such as quantum matrix inversion~\cite{PhysRevLett.103.150502, PhysRevLett.109.050505} or quantum machine learning~\cite{quantph.1307.0411, PhysRevLett.113.130503, Nature.10.631, quantph.1408.3106} that only require a number of queries polynomial in $n$, and those such as quantum searching~\cite{quantph.9605043} that require a number of queries super-polynomial in $n$. In the former case, as will be seen, the maximum qRAM gate error rate tolerated by the algorithms scales polynomially in $n$, and qRAM quantum error correction may not be required. In the latter case, which is the one that we concentrate on in this paper, the maximum tolerable qRAM error rate scales super-polynomially in $n$ and quantum error correction is needed.}

A conceptually simple physical implementation of a qRAM corresponds to a direct generalization of the fanout architecture used in classical RAMs. However, the number of faulty components that can be tolerated by the quantum architecture is of prime importance due to the difficulty in maintaining quantum coherence. This motivates searching for schemes with fewer faulty components.  A fundamental assumption of the qRAM architecture is that ``active" gates\footnote{The concept of ``active" gates introduced in \cite{PhysRevLett.100.160501, PhysRevA.78.052310} is somewhat unnatural when extended to quantum gates. At the physical level, a gate is considered active if it physically acts on its input. Since the qRAM may be in a superposition of querying many (or all) possible bit values in the memory, every gate may be in a superposition of being active or not. Implicitly, there is a physical process that is checking whether each gate is active, and then acting in that case, and such a process will not be perfect in practice. Translated into the circuit model, such gates may be modelled as controlled-gates, i.e. gates that act on its input provided that the control qubit is set to $\ket{1}$. Therefore, such a gate is considered ``active" if its control is set to $\ket{1}$ and ``non-active" otherwise.
\\ 
In practice, even non-active gates will be prone to errors. The implicit assumption is that these errors are much smaller than the errors in active gates, and the focus of the bucket brigade models is to reduce the impact of the higher order errors found in the active part of a gate.}
 are the only ones with significant errors.

In this paper we investigate the bucket brigade qRAM proposal introduced in \cite{PhysRevLett.100.160501, PhysRevA.78.052310}. Assuming one requires a constant error probability for the oracle query, then with the bucket brigade error model it suffices to have an error rate that is on the order of $\OC(1/n^2)$. In the bucket brigade model, one assumes that each computational path only contains $\OC(n)$ components that are faulty, and that a total of $\OC(n^2)$ faulty operations are performed. One can argue that it is optimistic to assume that the so-called ``non-active" components will be completely error-free. And, one could counter-argue that the error rates will be much lower, and thus ignored for problem instances of appropriate size. For the purposes of this article, we set aside these concerns and accept the premise of there only being $\OC(n)$ faulty components.

In contrast to such a qRAM, if one just used a regular fanout circuit for the lookup, with no error correction, one would need
to maintain quantum coherence over an exponential number of components~\cite{PhysRevA.78.052310}. In order to achieve a constant error rate for the query in this case, one would need to implement a fault-tolerant version of the look-up circuit, which would normally incur an overhead that is polynomial in $n$.
One advantage of bucket brigade qRAM is thus to bypass the poly-log overhead of fault tolerant quantum error correction needed to achieve a constant error rate for a look-up. Such an error rate would be sufficient if the qRAM is used in an algorithm making a constant number of queries, for example, for certain state generation algorithms \cite{Mosca:01, quantph.0208112}. In general, for an algorithm with inverse polynomially many queries, it would suffice to reduce the query error rate to be inverse polynomial in $n$, e.g. \cite{quantph.1307.0411, Childs:2003:EAS:780542.780552}.

In this article, we firstly shed doubt on the usefulness of a qRAM that provides queries with constant probability of error, when used with algorithms \add{such as quantum searching} that make super-polynomially many oracle queries. As an aside, we note that if the imperfect query operation is assumed to be unitary, and if one can apply the inverse of this imperfect query, then one can apply simple amplification methods to achieve queries with arbitrarily small error $\delta$ using a number of repetitions that is proportional to $\log(1/\delta)$. 
It was shown that this logarithmic overhead is not necessary for quantum searching \cite{Hoyer2003} and other problems \cite{quantph.0309220}. However, there is no reason to expect the errors in a realistic qRAM to behave this way, and in this article we consider incoherent errors.

We first show that a very simple model of incoherent physical errors induces an overall query error similar to the one described by Regev and Schiff~\cite{Regev:2008:IQS:1427895.1427975}. Consequently, a qRAM that produces queries with constant error will not permit the quadratic speed-up in Grover's search algorithm~\cite{quantph.9605043} or any other quantum search algorithm one might design. We show that one cannot escape achieving an error rate that is super-polynomially small. We conjecture that this error model nullifies the asymptotic speed-ups of other quantum query algorithms as well, and leave as open questions the extension of this result to other important query problems.

This negative result implies the need for some means of error reduction for the qRAM, with a look-up error rate exponential in $n$.  For consistency we assume a physical error rate that is inverse polynomial in $n$, the logarithm of the size of the database.  
We thus explore a natural approach, using quantum error correcting codes, to provide this error reduction, and argue that the apparent advantage of qRAM disappears in this case; in principle, one can make the error rate arbitrarily small, however the advantage of a small number of activated gates in the bucket brigade architecture appears to be lost when active error correction has to be performed on each gate. 
The main motivation for the quantum bucket brigade approach over a straightforward binary-tree approach is that the equivalent of the active gates are the only gates prone to error, and thus an inverse polynomial in $n$ error rate suffices in order to achieve an overall constant error per qRAM look-up.

The remainder of this paper is organized as follows. In Sec.~\ref{sct:qRAM Architecures} we describe the bucket brigade qRAM architecture and prove that for the Regev and Schiff model~\cite{Regev:2008:IQS:1427895.1427975} the error rate per gate must scale as inverse polynomial in the size of the database. In Sec.~\ref{sct:Realistic_Errors} we develop and analyze a simple error model that provides intuition for the overall behaviour of the memory with realistic noisy environments. In Sec.~\ref{sct:Circuit}, in order to discuss approaches for introducing quantum error correction inside the qRAM architecture, we introduce a circuit model for the bucket brigade architecture. We then argue in Sec.~\ref{sct:Error_Correction} that a fault-tolerant bucket brigade qRAM loses the advantage of small number of active components. Finally, in Sec.~\ref{sct:Conclusions} we conclude
and present some open problems and directions for future research.

%%%% qRAM Architectures %%%%

\section{Quantum RAM Architectures}\label{sct:qRAM Architecures} 
\begin{figure}[!htbp] 
\begin{center}
\includegraphics[scale=0.8]{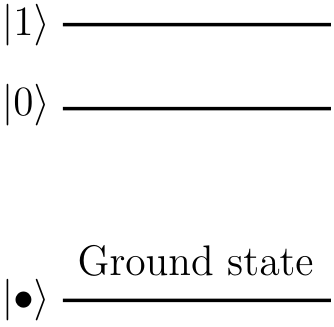} 
\end{center}
\caption{Representation of the energy levels of a qutrit used at the nodes of the routing binary tree.  The states $\ket{0}$ and $\ket{1}$ form a metastable subspace since the energy difference between the states is required to be much smaller than the difference between the ground state~$\ket{\bullet}$ and~$\ket{0}$.} 
\label{fgr:1} 
\end{figure}

In~\cite{PhysRevLett.100.160501, PhysRevA.78.052310}, Giovanetti \textit{et al.} proposed a quantum bucket brigade
addressing scheme requiring only $\OC(n)$ activations per memory call. The nodes of the routing binary tree are three level quantum systems (qutrits), with an energy spectrum schematically depicted
in Fig.~\ref{fgr:1}.

\begin{figure}
\centering
\includegraphics[scale=0.75]{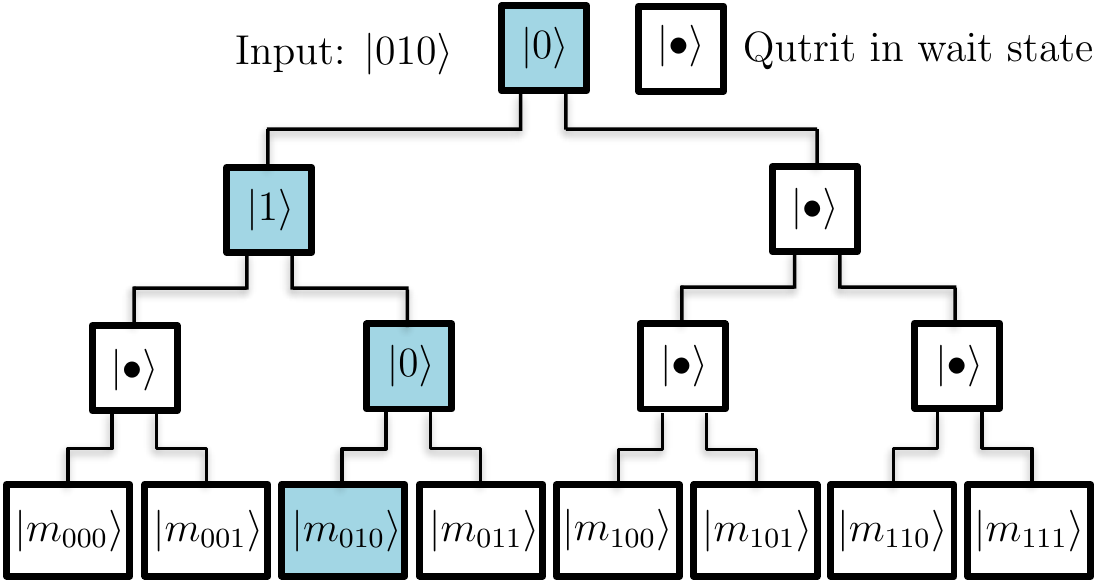}
\caption{(Color online) Bucket brigade scheme for a qRAM with 8 memory locations. The address register is $\ket{010}$, corresponding to the memory location $m_{010}$. The path $0\rightarrow1\rightarrow0$ is established by sequentially introducing the address qubits $\ket{010}$ into the root of the binary tree.} 
\label{fgr:2} 
\end{figure}

The $2^n$ qutrits at the nodes of the binary tree are initially prepared in the ground state $\ket{\bullet}$, named the ``wait" state, and the
memory address is specified by the $n$-qubit state $\ket{a_0a_1\ldots a_{n-1}}$.
At time $t_0$, the address qubit $\ket{a_0}$ is input
at the root of the tree and it interacts with the qutrit at node~$0$ changing its state from $\ket{\bullet}$ to $\ket{a_0}$. The states $\{\ket{0},\ket{1}\}$ of the node qutrit are coupled
to two spatial directions (paths), right and left respectively. The role of the coupling is
to route the  following incoming address photon along the correct 
path of the binary routing tree.
At time $t_1$, the subsequent address qubit $\ket{a_1}$ is input at the root of the tree. 
The address qubit $\ket{a_1}$ interacts with the qutrit at node $0$ and is physically routed down the left or right path of the tree depending upon the state $\ket{a_0}$ of node $0$. Consequently it changes the state of the corresponding node at level $1$ to $\ket{a_1}$. The process continues until all the remaining address qubits are sent through the tree, with the $k$-th address qubit changing the state of the node at the $k$-th level from $\ket{\bullet}$ to $\ket{a_k}$. After
$\OC(n^2)$ time steps\footnote{The $k$-th address qubit interacts with the first $k-1$ routing nodes, followed by a single interaction with the corresponding node at the $k$-th level. Considering each interaction takes a single time step, the $k$-th address qubit changes the state of the corresponding node at the $k$-th level after $k$ time steps. Considering there are a total of $n$ address qubits, the overall time required is $\OC(n^2).$}, a routing path is assigned from the root of the tree to the
desired memory location, with only $n$ nodes in the path (one node per level) having a state different from $\ket{\bullet}$.
A bucket brigade routing scheme for an $2^3$-address qRAM is schematically depicted in Fig.~\ref{fgr:2}.
The proposed physical implementation of bucket brigade in \cite{PhysRevA.78.052310} uses atoms in a cavity as routing nodes and polarization photon states as addressing qubits. 

In~\cite{PhysRevA.78.052310}, the authors claim that the bucket brigade scheme is coherent as long as the error per gate, $\varepsilon$ scales as $\OC(1/n^2)$. For this error scaling, as $n$ increases, the \emph{overall} error rate of the qRAM oracle asymptotically approaches a constant. Although constant or polynomial error rates suffice for some quantum algorithms \cite{quantph.1307.0411, Childs:2003:EAS:780542.780552}, such error rates are not favourable for some other important quantum algorithms. For example, Regev and Schiff~\cite{Regev:2008:IQS:1427895.1427975} showed that the quadratic speed-up in Grover's searching algorithm vanishes when using oracles with a constant error rate. 
Namely, in order to regain the quadratic speed-up, the error rate per oracle call should scale no worse than $
\OC(2^{-n/2})$ (therefore the error rate not only needs to be non-constant but it must vanish at a fast enough rate with increasing~$n$). 

In the next few sections, we construct a simple model of bucket brigade qRAM with errors and show in Appendix~\ref{apdxA} that Regev and Schiff error model~\cite{Regev:2008:IQS:1427895.1427975} resembles the model we construct. Based on this resemblance and assuming $\OC(n^2)$ faulty operations per memory call, we conjecture that in order to implement the qRAM for quantum searching, the overall error rate per memory call has to be in $\OC(2^{-n/2})$. In fact, for this to hold, the error rate per gate $\varepsilon$ should 
decrease faster than $1/f(n)$, where $f(n)\in \omega(2^{n/2})$. Thus $\varepsilon$ has to be in $o(2^{-n/2})$ and hence much smaller than $\OC(1/n^2)$, since the overall error rate per memory call must scale as
\begin{equation}\label{eqn:2}
1-\left(1-\frac{1}{f(n)}\right)^{\OC(n^2)}\in \Omega\left(\frac{n^2}{f(n)}\right),
\end{equation}
and in order to satisfy
\begin{equation}\label{eqn:3}
1-\left(1-\frac{1}{f(n)}\right)^{\OC(n^2)} \in \OC(2^{-n/2}),
\end{equation}
it is required that
\begin{align}\label{eqn:4}
f(n)\in\Omega&\left(n^22^{n/2}\right) \implies \frac{1}{f(n)}\in\OC\left(\frac{1}{n^22^{n/2}}\right)\notag\\
&\implies  \varepsilon \in o(2^{-n/2}).
\end{align}

Recently Hong et al. \cite{PhysRevA.86.010306} proposed a bucket brigade qRAM scheme in which the number of time steps required per memory call is reduced from $\OC(n^2)$ to $\OC(n)$. While this reduction decreases the overall error rate, the error rate per gate $\varepsilon$ must still be in $o(2^{-n/2})$.

The need for super-polynomially small (in $n$) error rate per gate for real world applications motivates a more thorough analysis of the bucket brigade qRAM scheme and the need for quantum error correction, these topics being the subject of the following sections.

%%%% Error Analysis %%%%
\section{Errors Analysis\label{sct:Realistic_Errors}}
%Start writing your part here without using the ``section" keyword. The output
%of only your section is obtain by compiling main.tex in the current folder.
%Uses qram.bib in the main folder as the bibtex file.
In this section we introduce a simple toy error model for the physical implementation
proposed in \cite{PhysRevA.78.052310}, in which the qutrits are implemented 
by trapped atoms in cavities. The address qubits are implemented by photons that
propagate along the network of cavities, and excite the corresponding qutrit 
to either of the states $\ket{0}$ or $\ket{1}$, depending
on their polarization. In this way, the incoming address
photons create a ``path" through the binary tree of cavities, leading to
the desired memory location. The readout is performed by injecting a
``bus" qubit (photon) at the root of the tree that interacts with the desired memory location,
copies its value (the states stored by the memory are $\ket{0}$ or $\ket{1}$, and 
not any superposition), and finally is sent back along the routing tree exiting through the
root with the corresponding memory location content.
For more details about the physical model an interested reader
is referred to \cite{PhysRevA.78.052310}.

\subsection{Toy Error Model}\label{sbsct:Toy Error Model}
In the following we assume that the only source of errors in the above model
is due to random flips between the states $\ket{0}$ and $\ket{1}$ of the qutrit.
We assume a typical symmetric bit-flip error, in which 
at each time step the state $\ket{j}$ can either flip to
 $\ket{j\oplus 1}$ with probability $\varepsilon$ or remain unchanged with probability $1-\varepsilon$.
The motivation for considering this error model
 is that, since the states $\{\ket{0},\ket{1}\}$ are close together in the energy
 spectrum, significantly less energy is required to cause a flip between them,
 hence such flips are more likely to occur.
 In reality, there may be other sources of errors such as coupling errors, decaying 
of excited qutrit states to the ground state, loss of photons during the routing process and so on. 
However, our toy model illustrates the effects of an error that would naturally occur in a realistic physical realization of a qRAM. There is no reason to expect these other sources of errors would help matters (otherwise, one could seek to deliberately introduce or simulate such errors).
 
 It is not hard to observe that any error
 in the routing process can propagate through the tree resulting in various
 possibilities. Considering all possible errors in such a model, the possible
 paths that the bus photon could take in the final step termed as \emph{right path},
\emph{wrong path} and \emph{no-path}, respectively. For convenience, we further assume the operations used to un-compute the path information encoded in the qRAM are error-free.

\emph{1) Right path} -- This scenario occurs when no flips (errors) arise
			during the routing process.
			In this ideal scenario, the bus reaches the correct location in the
			qRAM as specified by the input address. Fig.~\ref{fgr:2}
			depicts an example of a \emph{right path} given an input address $\ket{010}$.
			
 To compute the probability $p_{rp}$ of such an event, we require that no bit flip occurs at each of the 
 $j$ levels. Taking the intersection of such events for
all $n-1$ levels of the binary tree gives the probability of the \emph{right path}

\begin{align} 
 p_{rp}&=\prod\limits_{j=0}^{n-1} (1-\varepsilon)^{n-j}=(1-\varepsilon)^{\sum\limits_{j=0}^{n-1} (n-j)}\notag \\ 
 &=(1-\varepsilon)^{n(n+1)/2}.
\label{eqn:5} 
\end{align}

\begin{figure}[!ht] 
\begin{center}
\includegraphics[scale=0.75]{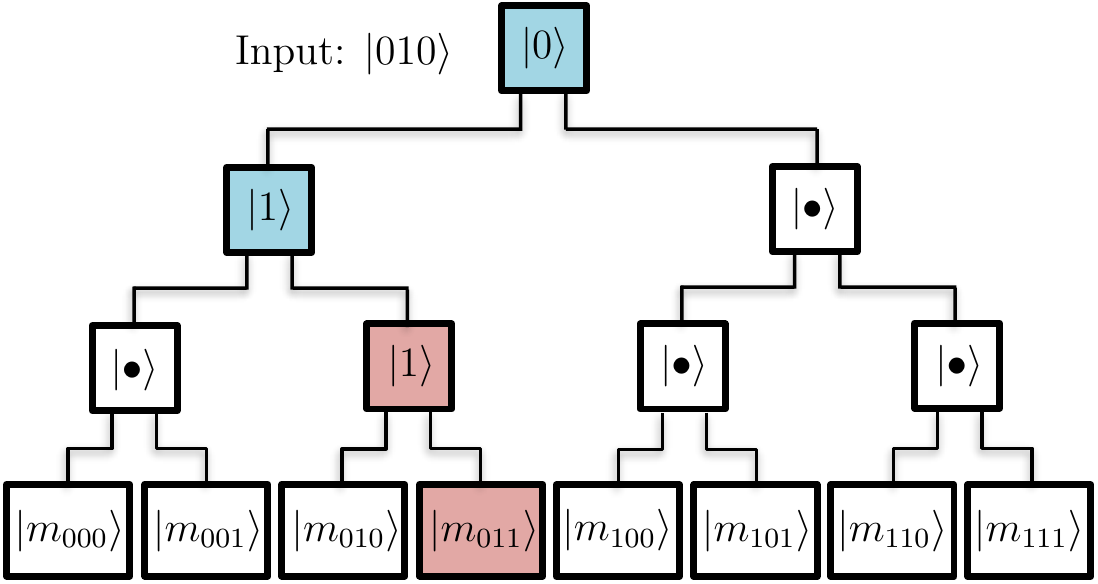} 
\end{center}
\caption{(Color online) Example of a \emph{wrong path} produced by an error at the third time step, given the address
		$\ket{010}$.} 
\label{fgr:3} 
\end{figure}

\begin{figure}[!ht] 
\begin{center}
\includegraphics[scale=0.75]{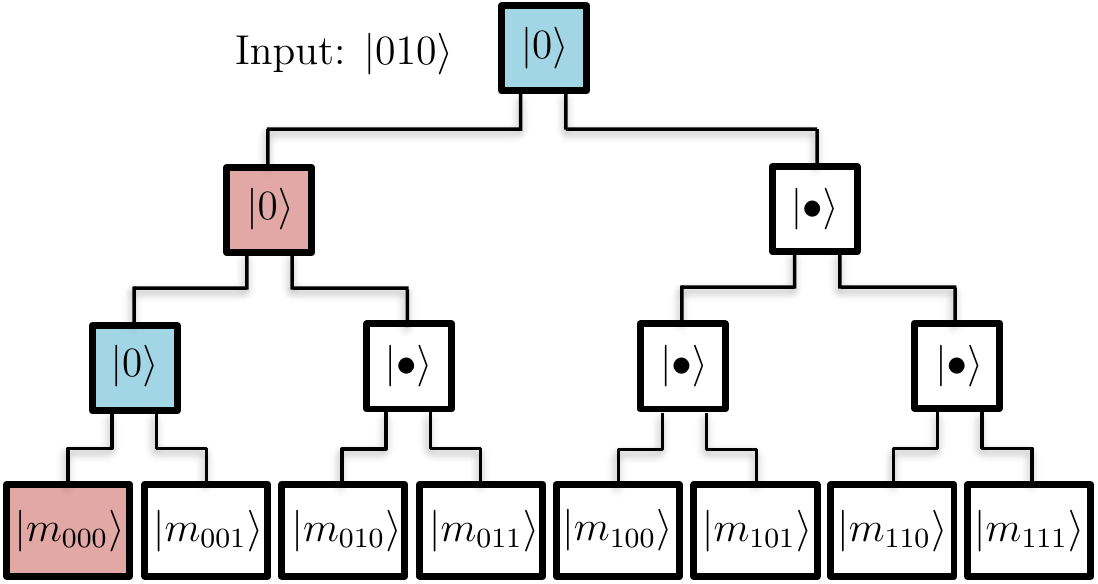} 
\end{center}
\caption{(Color online) Example of a \emph{wrong path} produced by an error at the second time step, given the address
		$\ket{010}$.} 
\label{fgr:4} 
\end{figure}

\emph{2) Wrong path} -- This error refers to the cases wherein the 
	the bus reaches \emph{any  other} location in the qRAM other than the
	location corresponding to the input address. A \emph{wrong path} error occurs at level
	$i$ if the state $\ket{j}$ of the active routing qutrit at level $i$ flips to $\ket{j\oplus 1}$ and no  
	other errors occur subsequently (at later time steps). 
	The scenario where another error occurs at a later time step in the levels preceding to the $j$-th
	level leads to a \emph{no-path} error which we discuss later.
	The following two figures
	illustrate two possible \emph{wrong path}s for the input  address $\ket{010}$. In
	Fig.~\ref{fgr:3}, the error is assumed to occur in the third time
	step, due to which the bus accesses the wrong location corresponding to $\ket{011}$.
	In Fig.~\ref{fgr:4}, the error is assumed to occur in the second time
	step, with the bus wrongly accessing the location corresponding to
	$\ket{000}$.

In order to calculate the probability of a \textit{wrong path} occurring, we consider the probability of any path occurring, regardless of whether it is the \textit{right} or~\textit{wrong} path, we denote this probability by~$p_{path}$. Suppose the state~$\ket{\psi_j}$ is being routed down the qRAM circuit to the $j$-th level. If any of the $(j-2)$ first routing nodes have flipped then the state will be routed down an unexpected branch and will not excite the~$j$-th level of the tree, resulting in a \textit{no-path}. The probability of success at this given time step is therefore~$(1-\varepsilon)^{j-1}$, where~$\varepsilon$ is the probability of a node flipping, recall we must include the level-0 root node here. This can only for levels~2 and above. The overall probability of success is therefore the product of each of the individual probabilities of success at each time step (including the time step to send the bus qubit down the tree to recover the information stored in the RAM). This probability is given by:

%
%We have not illustrated the possibility of multiple errors creating a \emph{wrong path}. 
%This could happen as long as a \emph{no-path} is not created. 
%Note that 
%if at a given time instant a qutrit in the $j$-{th} level of the tree is
%activated and if at any subsequent time an error occurs on a level before the $(j-1)$-th level, then the resulting
%error is a \emph{no-path} error. An error only on the $(j-1)$-{th} level results
%in a \emph{wrong path} and cannot create a \emph{no-path}. The probability of the bus
%reaching \emph{any} location on the qRAM (irrespective of a \emph{right path} or wrong
%path) is

\begin{align} 
p_{path}&=p_{wp}+p_{rp}=\prod\limits_{j=2}^{n}
	(1-\varepsilon)^{j-1}\notag\\
	&=(1-\varepsilon)^{\sum\limits_{j=2}^{n}(j-1)}=(1-\varepsilon)^{n(n-1)/2}.
\label{eqn:6}
\end{align}
%Note that the product is up to the second last level of the tree $n-2$, since an error in the last level
%can not create a \emph{no-path}.

As we computed before the probability of a \emph{right path} $p_{rp}$ in Eq.~\eqref{eqn:5},
the probability of a \emph{wrong path} is then 
\begin{align}
\label{eqn:7} 
p_{wp}&=p_{path}-{p_{rp}}\notag\\
&=(1-\varepsilon)^{n(n-1)/2}-(1-\varepsilon)^{n(n+1)/2}.
\end{align}

\begin{figure}[!htbp] 
\begin{center}
\includegraphics[scale=0.75]{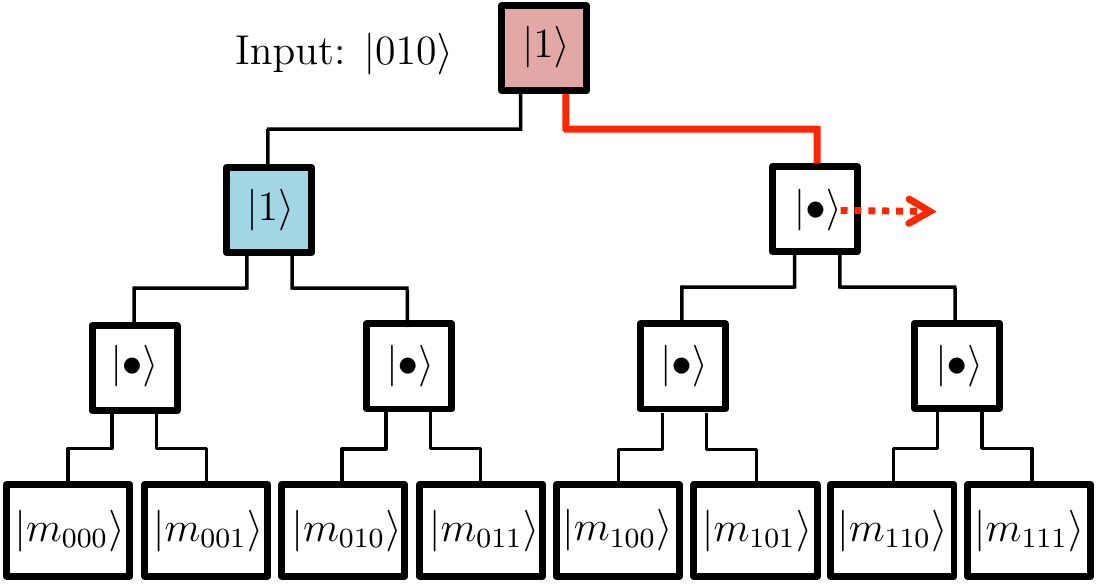} 
\end{center}
\caption{(Color online) Example of a \emph{no-path} given the address $\ket{010}$}
\label{fgr:5} 
\end{figure}

\emph{3) No-path} -- This error refers to the scenario where the bus never
	reaches {\it any} location of the qRAM.
	Such an error arises when a bit flip error occurs in levels $0$ to $n - 3$.
	The smallest such tree where this error can occur is therefore a
	three-level tree (corresponding to a qRAM with $2^3$ memory cells), as shown in
	Fig.~\ref{fgr:5}. The difference between a \emph{wrong path} and 
	a \emph{no-path} is that, in the latter, the bus photon does not reach
	the memory address, hence does not read any information,
	whereas in the former scenario the bus reaches the wrong address
	in the qRAM and after the un-computing stage, the bus contains the information of 
	\emph{some} particular address in the qRAM. 

	We present an example of a \emph{no-path} error in Fig.~\ref{fgr:5}, for 
	an input address $010$.
	At the first time instant, the first address photon
	(i.e. $\ket{0}$) activates the switch (qutrit) in the first layer of the
	tree. At the second time instant, the address photon $\ket{1}$ interacts with the
	switch in the first layer, now in state $\ket{0}$, to decide the direction in which
	it has to be routed. Assuming no error during
	the second time step, the second address photon is correctly routed to the left path.
	Assume now that at the third time instant, a flip error occurs on the root
	qutrit, which flips its state from $\ket{0}$ to $\ket{1}$. 
	The third address photon would then be incorrectly routed to the path on the right.
	As it can be seen from Fig.~\ref{fgr:5}, at the third time instant
	there are two activated switches in the second level. 
	The readout bus photon can no longer reach any of the memory locations,
	and will be lost in the second level of routing tree.
 
 The probability of a \emph{no-path} event is simply
 \begin{equation}\label{eqn:8}
 p_{np}=1-p_{wp}-p_{rp}=1-(1-\varepsilon)^{n(n-1)/2}.
 \end{equation}

If the qRAM is used to implement a quantum oracle $O$, then $O$ will be faulty, 
with an error model described by 
\begin{equation}\label{eqn:9}
\rho \stackrel{O}{\rightarrow } p_{rp} \hat O\rho \hat O^{\dagger} + p_{wp} \EC_{wp}(
	\rho)+ p_{np}\EC_{np}(\rho),
\end{equation}
with $\hat O$ denoting a perfect oracle.
Here $\EC_{wp}(\cdot)$ and $\EC_{np}(\cdot)$ are error channels that corresponds to the
\emph{wrong path} and \emph{no-path} errors, respectively. 

Our error model Eq.~\eqref{eqn:9} is less optimistic than the one of Regev and Schiff~\cite{Regev:2008:IQS:1427895.1427975} of the form $\rho\stackrel{O}{\rightarrow}(1-p)\hat O\rho \hat O^\dagger+p\rho$. The main difference is that the latter does not mix the amplitudes of the initial starting superposition state in Grover's search algorithm, whereas our model decoheres the system much faster due to the non-trivial errors $\EC_{wp}$ and $\EC_{np}$. Although we do not have a proof that the quantum query complexity of our model cannot be less than the one considered in \cite{Regev:2008:IQS:1427895.1427975} (i.e. linear in $N$), we conjecture (based on a formal proof for a similar decoherence model, see Appendix~\ref{apdxA}) that this is indeed the case. 

\subsection{Asymptotic Behaviour\label{sbsct:Error_Analysis}} 
\begin{figure}[!htbp]
\includegraphics[scale=0.86]{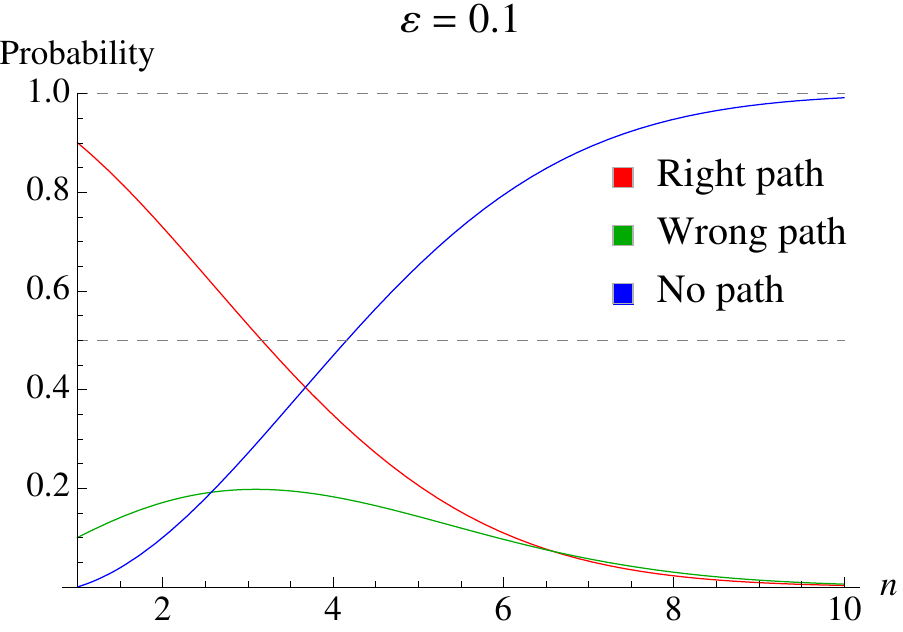}
\caption{(Color online) Comparison of errors for fixed $\varepsilon$ as a function of $n$.}
\label{fgr:6} 
\end{figure}

\begin{figure}[!htbp]
\includegraphics[scale=0.86]{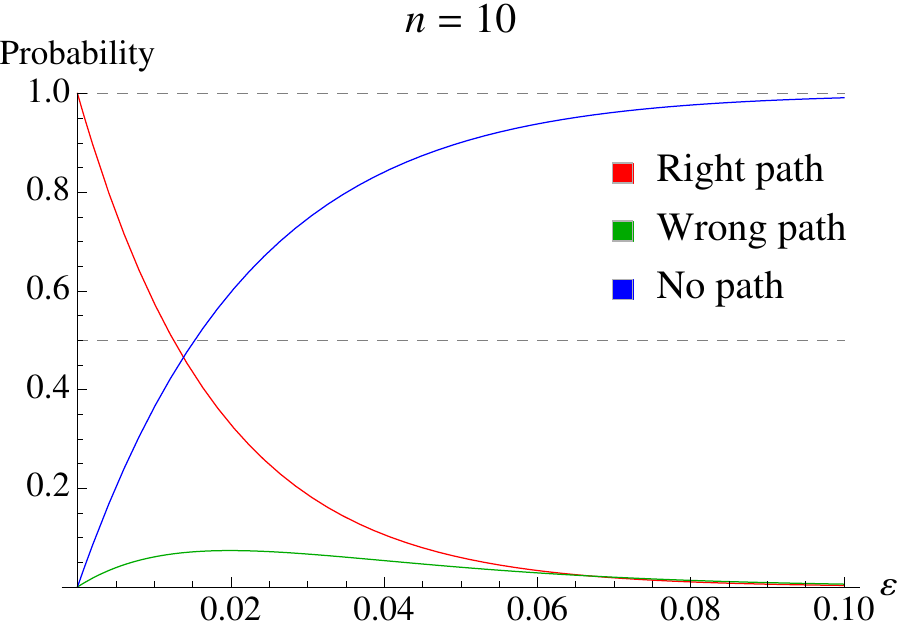}
\caption{(Color online) Comparison of errors for fixed $n$ as a function of $\varepsilon$.}
\label{fgr:7} 
\end{figure}

\begin{figure}[!htbp]
\includegraphics[scale=0.86]{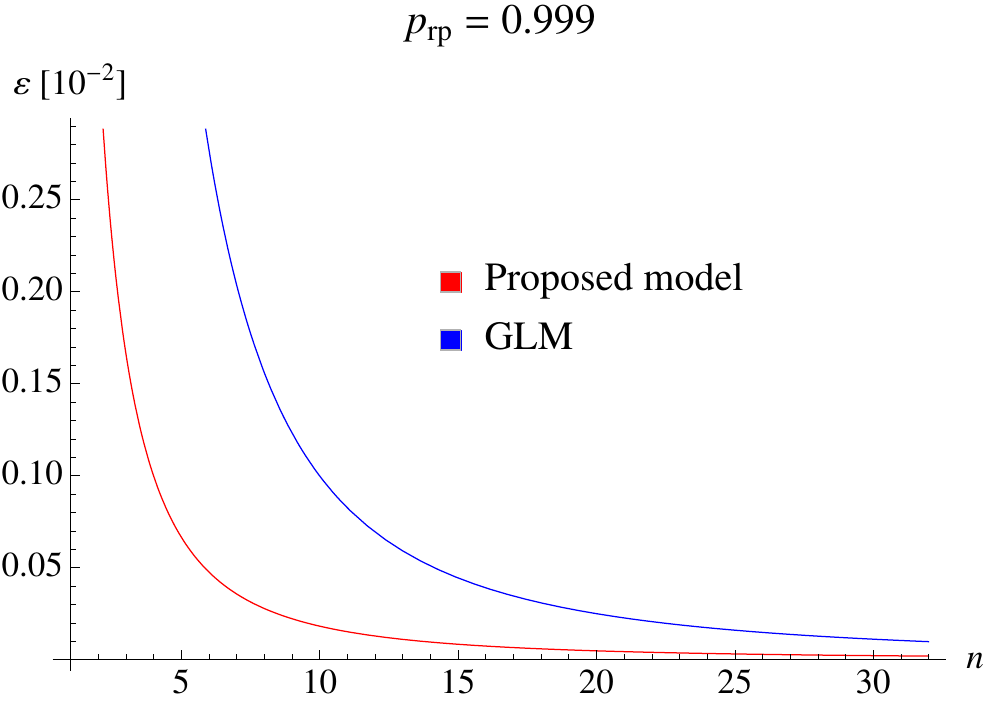}
 \caption{(Color online) Required $\varepsilon$ (in dimensionless units of $10^{-2}$)
	as a function of $n$, for a fixed circuit fidelity. GLM denotes the model proposed in \cite{PhysRevA.78.052310}.} 
\label{fgr:8}
\end{figure}

In Figs.~\ref{fgr:6}, ~\ref{fgr:7} and ~\ref{fgr:8} we analyze the
probabilities of the three types of errors discussed in the previous subsection. 
The parameters of interest are the
error probability per gate, denoted by $\varepsilon$, the overall fidelity of the
addressing circuit (i.e. the probability of a \emph{right-path}), denoted by
$p_{rp}$, and the number of levels in the qRAM addressing binary tree denoted
by $n$ (corresponding to $2^n$ memory locations). 

For a fixed $\varepsilon$, we see that the \emph{no-path} factor dominates in the error model, asymptotically with $n$, as depicted in
Fig.~\ref{fgr:6}.  

For a fixed $n$, again the \emph{no-path} error dominates when the error per gate
$\varepsilon$ becomes large, see Fig.~\ref{fgr:7}.  

Finally, for a fixed desired overall fidelity $p_{rp}$, the maximum
allowed error probability per gate $\varepsilon$ to achieve the overall fidelity $p_{rp}$ 
decays exponentially as a function of $n$, as plotted in Fig.~\ref{fgr:8}.  
From Fig.~\ref{fgr:8} it can be seen that, the error rate per gate of $\OC(1/n^2)$ (blue line in Fig.~\ref{fgr:8}) as considered in Giovannetti et al. \cite{PhysRevA.78.052310} is more optimistic than our error rate $\varepsilon(n)$ (red line in Fig.~\ref{fgr:8})

%In comparison, Giovannetti et al. \cite{PhysRevA.78.052310} considered an error rate per gate that scales as $\OC(1/n^2)$, a more optimistic  model than ours, as can be visually inspected in Fig.~\ref{fgr:8} by comparing the graphs of $1/n^2$ (blue line if Fig.~\ref{fgr:8}) and $\varepsilon(n)$ (red line in Fig.~\ref{fgr:8}). 
For larger output fidelity
$p_{rp}$, $\varepsilon(n)$ will always  be bounded above by
$1/n^2$, with the gap between the two increasing as $p_{rp}$ approaches towards 1. Asymptotically in $n$, the two graphs converge towards zero. 

Simply, the difference between our error $\varepsilon(n)$ and the one in
\cite{PhysRevA.78.052310} can best be understood by investigating
the series expansion
\begin{align}
\label{eqn:10}
	p_{rp}=(1-\varepsilon)^{n^2}=1+2\log(1-\varepsilon)\frac{1}{n(n+1)}+\OC(\frac{1}{n^4}).
\end{align} 
In \cite{PhysRevA.78.052310} the authors considered only the first order $1/n^2$ as a desirable error rate
per gate. However, when the output fidelity $p_{rp}$ approaches 1, this approximation is no longer accurate, and higher order terms are important. As mentioned at the end of 
Sec.~\ref{sct:qRAM Architecures}, inverse polynomial error rates are not good enough
in implementing Grover's search with a qRAM-based oracle. In fact, overall error rates of at most 
$\OC(2^{-n/2})$ are essential. 

The dominant \emph{no-path} error term poses a fundamental implementation problem, due to lack of oracle information, similar (see Appendix~\ref{apdxA}) to the noise model investigated by Regev and Schiff~\cite{Regev:2008:IQS:1427895.1427975}. If in the future, qRAM designs could be constructed without the presence of such a \emph{no-path} term (i.e. with \emph{only wrong-path} noise), one can attempt error correction to efficiently reduce the error rate. We demonstrate in Appendix~\ref{apdxB} a possible error correction scheme for a simplified \emph{wrong-path} term governed by bit-flip channels, then show however that the scheme is neither applicable to our error model nor to the Regev and Schiff error model~\cite{Regev:2008:IQS:1427895.1427975}.

%%%% Circuit model %%%%
\section{Circuit Model\label{sct:Circuit}}
\begin{figure*}[!ht]
 \centering
\includegraphics[scale=0.85]{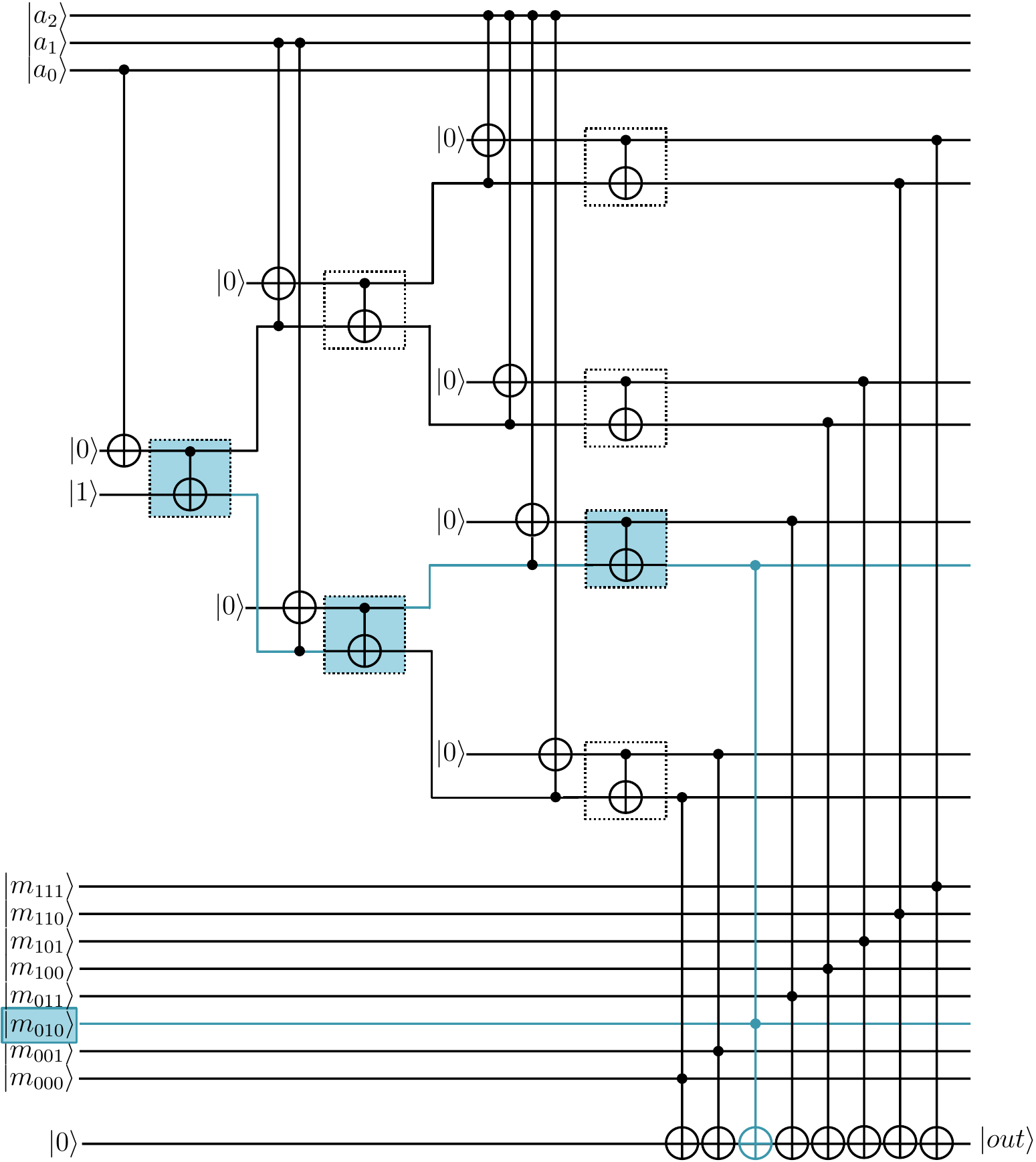} 
\caption{(Color online) Circuit for bucket brigade qRAM. Nodes to the left of the memory cell are
\emph{routing nodes}. The dashed squares 
represents the memory locations. The first layer of nodes immediately to the right of the memory
are the \emph{coupling nodes}. Finally, the nodes on the right are the \emph{read out nodes}. 
A possible input is e.g. $\ket{a_0a_1a_2}=\ket{010}$, for which the circuit reads the memory location $m_{010}$. 
The path leading to the location $m_{010}$ is represented in blue colour, and the active routing and readout nodes are highlighted.
One could more closely mimic the physical flow of information in the bucket brigade qRAM by adding an additional qubit at each node in the binary tree we see in the diagram. Then, for $k \in \{0,\ldots,n-1 \}$, we add an initial controlled-NOT gate to copy $a_k$ to the root node, followed by a series of $\OC(2^k)$ controlled-SWAPs that will bring the value of $a_k$ to the unique node in level $k$ defined by the bits $a_0, a_1, \ldots, a_{k-1}$. While this adds exponentially many gates, it does not change the overall gate complexity, and these additional gates only add $\OC(k)$ to the depth of the circuit. This also illustrates that the exponential depth implicit in the circuit we describe in the diagram can easily be reduced to polynomial depth by further mimicking the ideas presented in the qRAM proposal. We leave the circuit diagram in this simpler form, since it does not affect our arguments in Sections~\ref{sct:Realistic_Errors} and~\ref{sct:Error_Correction}.
} 
\label{fgr:9}
\end{figure*}

To facilitate the discussion of error correction, in this section we reformulate the physical model of bucket brigade qRAM in \citep{PhysRevA.78.052310} as a quantum circuit.
In Fig.~\ref{fgr:9} we present a possible circuit description for an $N=2^3$ qubit bucket brigade qRAM, in which
the memory contains only states in the computational basis $\{\ket{0}, \ket{1}\}$. Our circuit
is immediately extendable to $N=2^n$ and closely simulates the physical model\footnote{Further modifications could be made to more closely mimic this model, and are discussed in the caption of Figure~\ref{fgr:9}.}
proposed in \cite{PhysRevA.78.052310}. 

The circuit description of the bucket brigade addressing scheme accounts for the temporal aspects of the bucket brigade scheme. Namely, since the address qubits are introduced into the binary tree architecture sequentially, the circuit description should respect this ordering. The input to the circuit are the address qubits $\ket{a_0 a_1 \hdots a_n}$. The circuit resembles a binary tree composed of~$2^n-1$ routing nodes, $2^n$~memory cells and $2^n$~readout nodes that perform the inverse operations of the routing circuit, used to decouple the qRAM from the address qubits. Additionally, a bus qubit is introduced that interacts with the memory nodes to extract the information stored in the appropriate memory location. It is worth noting that this bus qubit as described may not be physically realistic since it may interact with all the bits in the qRAM. We leave it as such, for simplicity. In practice, if such a non-local qubit is not feasible, one may either work with a phase oracle (as described in Ch.~8 of \cite{KLM07}), or one may use a binary-tree circuit to bring the result of the qRAM look-up to a specific qubit that will be accessed by the quantum algorithm that performs the look-up.

The address qubit~$\ket{a_0}$ is used to activate the appropriate branch at the first level of the routing. The address qubit is coupled via a CNOT to an ancillary state prepared in the state~$\ket{0}$. This qubit then serves as one of the input qubits along with an additional qubit prepared in the~$\ket{1}$ state for the routing node (a CNOT gate with the first qubit as control). Depending on the state of the address qubit, the resulting two-qubit output of the routing node will have a single excited qubit in the $\ket{1}$~state, which we shall call the activated qubit. The activated branch of the tree governs the routing of the subsequent interactions with the address qubits, playing the role of the routing atom in the case of the bucket brigade outlined in~\cite{PhysRevA.78.052310}.

The two qubits at the exit of the level-0 routing node serve as inputs to the second register of the two level-1 routing nodes. These qubits control which of the routing nodes are activated at the next level of the qRAM binary tree architecture. Namely, the qubit that is excited in the~$\ket{1}$ state allows for the coupling between the address qubit and an introduced~$\ket{0}$ state ancilla via a Toffoli gate. Therefore the input to the active routing node is either~$\ket{01}$ or~$\ket{11}$ depending on the state of the address qubit~$\ket{a_1}$. Effectively, the routing operation given by a CNOT gate activates a branch of the tree. For the node that is non-active, the state at the output of the previous level is~$\ket{0}$, meaning the Toffoli is not activated and the resulting input and output state to the routing node remains~$\ket{00}$.
%Consider one such level-1 routing node, the qubit that comes from the output of the previous level along with the address qubit~$\ket{a_1}$ are used as the control qubits for a Toffoli gate, flipping the state of an ancilla qubit prepared in the~$\ket{0}$ state if both control qubits are in the state~$\ket{1}$. The qubit that comes from the output of the previous routing node will control whether the address qubit will have an effect on the ancilla state. Namely, if this qubit is in the state~$\ket{0}$ then this path of the qRAM will remain non-excited and the introduced ancilla state will remain in the state~$\ket{0}$, regardless of the state of the address qubit. Therefore, when these two qubits are inputted into the routing node composed of a CNOT gate, the output will remain in the $\ket{00}$ state, meaning that the output branches remain non-excited. However, if the qubit from the previous routing level is excited, then the interaction with the address qubit will occur, and as such the input to the routing node will be in the subspace spanned by the states~$\{ \ket{01}, \ket{11} \}$, with the resulting state depending on the state of the address qubit~$\ket{a_1}$. Therefore, the output of the routing node (composed of a CNOT) will be a state in the subspace~$\{ \ket{01}, \ket{10} \}$, thus only having a single excited branch, depending on the state of the address qubit. 
Therefore, after two routing node levels, the output of the routing qubits is composed of $2^2$~qubits, with only a single branch being excited depending on the state of the first two address qubits~$\ket{a_0 a_1}$. Therefore, this corresponds to an isometry:
\begin{align}
\ket{00}_{a_0 a_1} &\rightarrow \ket{0001} \nonumber \\
\ket{01}_{a_0 a_1} &\rightarrow \ket{0010} \nonumber \\
\ket{10}_{a_0 a_1} &\rightarrow \ket{0100} \nonumber \\
\ket{11}_{a_0 a_1} &\rightarrow \ket{1000},
\label{eqn:11}
\end{align}
where the excited output qubits in the~$\ket{1}$ state represent an active physical path for the subsequent qRAM operations.
This procedure is repeated for~$n$ levels, where at the $k$-th level there are $2^k$ Toffoli gates and routing nodes. The $2^k$ Toffoli gates are required to route of the address qubit~$\ket{a_k}$ through the previous~$k$ levels and the routing node establish the output states in order to route the subsequent address qubits. Since such a circuit performs the appropriate unitary mapping of the address qubits for all computational basis state inputs, by linearity it will extend to all superpositions of input address qubits. An example of the routing procedure for a three-qubit input address state~$\ket{010}$ is presented in Fig.~\ref{fgr:9}, where the blue highlighted nodes correspond to the activated nodes.

After the completion of the $n$~routing node levels, memory readout is performed. The reading is performed by introducing a bus qubit prepared in the~$\ket{0}$ state which is the target of $2^n$~Toffoli gates. Each of the $2^n$ qubits at the output of the qRAM routing nodes pair with one of the memory cells to serve as control qubits for the Toffoli gates. Since only a single output qubit from the routing scheme is activated, only a single Toffoli gate couples with a memory location to the bus qubit. The bus qubit is represented by the bottom qubit in Fig.~\ref{fgr:9} while the $2^3$~memory qubits are represented between the qRAM routing architecture and the bus qubit. 

Having completed the coupling of the address qubit, the state of the qRAM routing qubits must be decoupled from the address and bus qubits. Each of the gates from the routing circuit are performed in reverse order, which corresponds to performing the inverse unitary coupling transformation between the address qubits and the routing qubits. The resulting state couples the address qubits with the corresponding memory qubit and has decouples the routing qubits to their input ancillary states.

%%%% Error Correction %%%%
\section{Error Correction\label{sct:Error_Correction}} The results from
Sec.~\ref{sct:Realistic_Errors} motivate the
need for quantum error correction to be implemented at each node in order to
protect against errors that may cause detrimental faults in path information. 

\subsection{Imposing a quantum error correcting code} \label{sbsct:QECC} In
choosing a quantum error correcting code~(QECC) to protect the path information that is
stored in each node, it is essential to choose an encoding that can be
implemented fault-tolerantly, to allow for the generalization to large
computational systems. Moreover, the QECC should be
chosen such that it can naturally be incorporated from the quantum computer that
is accessing the qRAM. In order to analyze the desired error correction properties of a bucket brigade qRAM architecture we consider the circuit presented in Fig.~\ref{fgr:9}. The key gate components at each site are the CNOT and Toffoli 
gates. 

The most natural construction of a QECC that can
implement such operations with minimal overhead would be the 15-qubit
Reed-Muller code. The reason for choosing such a code would be that decomposing
the gate operations in the routing circuit as a sequence of CNOT and Toffoli
gates has the advantage that each of these gates can be implemented in a transversal manner.
Transversality is defined as the ability to implement a logical gate by applying physical
gates that have support at at most a single location per encoded codeblock: it is
the most natural way to guarantee fault-tolerance. However, if the quantum
computing device leads more naturally to another form of quantum error
correction encoding, methods such as state distillation or other schemes for
universal fault-tolerance can be used~\cite{PhysRevA.71.022316, PhysRevLett.111.090505, bombin2013optimal, PhysRevLett.112.010505, PhysRevLett.113.080501}.

% CSS codes and transversal CNOT gates
The focus of many fault-tolerant implementations are through the CSS code
construction~\cite{PhysRevLett.77.793, PhysRevA.54.1098, steane1996multiple}. A CSS quantum code is constructed using
two classical error correcting codes, each individually used to address $X$ and
$Z$~type errors. Given that any quantum error can be decomposed in terms of a
linear combination of Pauli operators, developing an error correcting code that
can address both types of errors will be sufficient for the construction of a~QECC. 

Let $\CC_X$ be a classical error correcting code of length~$n$ that has the
associated parity check matrix $H_X$, where each 0 in the parity check matrix of
the classical code is replaced by the two-dimensional identity matrix~$I$ and
each 1 in the classical parity check matrix is replaced by the Pauli~$X$
operator. Similarly, let $\CC_Z$ be a second classical error correcting code of
length~$n$ with an associated parity check matrix~$H_Z$, where each 1 in the
classical parity check matrix has been now replaced by the Pauli~$Z$ operator.
If $\CC_X^{\perp} \subseteq \CC_Z$ then by combining the stabilizers generated
from the parity check matrices of both codes, $H_X$ and $H_Z$, the resulting
stabilizer code forms a~QECC. The number of physical
qubits in the code is~$n$, the number of logical qubits is given by $k_X+k_Z-n$,
where $k_i$ is the number of logical states in the given classical code~$i$ and
the distance of the code is at least the minimum of the distance of the two
classical codes. One of the many appealing features of
the CSS code construction is the transversality of the CNOT gate, a feature of
the $X$~and $Z$~stabilizers being independent. A particular example of a CSS code is the 15-qubit Reed-Muller code mentioned above.

\subsection{Number of activations in a CSS code} \label{sbsct:ActivationsCSS}
In the implementation of Giovanetti~\emph{et al.}~\cite{PhysRevA.78.052310}, one
of the primary advantages is the number of gate activations that
are needed per level of the bucket brigade scheme. More simply, a CNOT~(Toffoli) gate in
their scheme is ``activated" only when the control qubit(s) is~(are) in
the state~$\ket{1}$. Since only one register is in such a state at a given level, the
total number of activations can thus be kept low. In a physical implementation, this is relevant as an activated path may represent the presence of a physical excitation without which no physical process occurs, therefore one can think of these non-activated gates as in fact being the identity operation. However, such an advantage vanishes when imposing a CSS code in order to protect from errors due to the
symmetry in the number of~$\ket{0}$ and~$\ket{1}$ states the logical states encoding the path information.

% Even number of activations of CNOT gates in CSS code construction
In the CSS code construction, two classical codes were taken to form a QECC. Therefore, given some codeword of the classical code~$c \in \CC_Z$, the
equivalent quantum state written out in computation basis~$\ket{c}$ must be
stabilized by the $Z$ generators of the code, by definition of being a codeword
of the classical code~$\CC_Z$. However, in order to be a logical state of the CSS code,
it must also be stabilized by the elements of the group generated by the
$X$~stabilizers. Therefore, the codestate will be the superposition of the
application of all $X$ stabilizers upon~$\ket{c}$, 
\begin{equation}
\label{eqn:12} 
\ket{c + \CC_Z^{\perp}} = \sum_{x \in \CC_Z^{\perp}}\ket{c+x} = \dfrac{1}{2^{n-k_X}}\prod_i (I^{\otimes n} + S_{X,i})\ket{c}, 
\end{equation} where $\{S_{X,i} \}$ 
are the generators of the $X$~stabilizer group, equivalently given by the rows of the parity
check matrix~$H_X$.

Consider the form of Eq.~\eqref{eqn:12}, given the state~$\ket{c}$ written
in the computational basis, the action of the operator~$(I^{\otimes n} +
S_{X,1})$ will be the equally weighted superposition of the state~$\ket{c}$ and
$S_{X,1}\ket{c}$, which will differ at the location where $S_{X,1}$ has a
Pauli~$X$ in its description. Therefore, at these locations half of the states
in the superposition will have a physical $\ket{0}$~state and half will have a
physical $\ket{1}$~state. Then acting upon the state with the
operator~$(I^{\otimes n} + S_{X,2})$ will have the same effect on all the states
in the superposition, with now an even number of physical~$\ket{0}$
and~$\ket{1}$ states occurring at location with Pauli~$X$ in $S_{X,2}$.
Repeating this for all X generators, any location with a~$X$ operator in one of
the stabilizers will necessarily have half of the states in the superposition in
each of the physical basis states. In order for the code to protect against any
arbitrary single qubit error, each physical qubit must be protected by at least
one X stabilizer operator with support the given location, otherwise it would be
vulnerable to a single~$Z$ error at this location. As such, all relevant CSS
codesstate will have an equal number of each of the physical basis states when
writing out the expansion of the state in the computational basis. 

In a physical implementation, such as that of Giovanetti~\emph{et al.}~\cite{PhysRevA.78.052310}, a qubit in the state~$\ket{1}$ represents an activated physical process, and as such the advantage of the bucket brigade scheme is that the number of such processes are kept low. However, due to the symmetry in the number of activations that must exist in both the logical ground and excited states, this advantage no longer exists when considering CSS~codes. More generally, non-symmetric codes, that is codes where the logical~$\ket{0}$ state and logical~$\ket{1}$ have a differing number of physical states in the excited state~$\ket{1}$, are not desirable for the purposes of error correction as they will be more susceptible to~$Z$ errors. The three-qubit repetition code is an extreme example of such a property.
% Comparison to physical implementation proposed by Giovanetti et at.

In principle, for the physical error model discussed in Section~\ref{sct:Realistic_Errors}, one can envision using the detection of a photon lost in the routing structure as a means to correct for~\textit{no-path} errors (see Figure~\ref{fgr:5}). However, detecting the exact node at which a photon was lost reveals path information about the state being read by the qRAM (since the previous node in the routing structure would have necessarily been activated by the address qubits) which leads to a loss of coherence in the system. Therefore, any photon detection has to identify the level at which the photon was lost, while not revealing exactly where. 
It is hard to envisage a practical means for experimentally realizing a photon detection with this property (for example, by somehow symmetrizing the loss of the photon across the exponentially many nodes at a given level).
Furthermore, even if this is achieved, one still faces the problem that the lost photon contained path information.
Thus, destroying the photon with this path information is equivalent to a dephasing error leading to a further loss of coherence.

%While such a restriction poses experimental difficulties, one could envision an experimental setup that symmetrizes the loss of a photon at a given level using a form of time binning. However, the scheme is further complicated by the following %theoretical 
%challenge: the lost photon contains path information itself, therefore it is important that upon its detection no information about its state is revealed. In order to overcome such an experimental obstacle, one could measure and discard the information of the photon's state. However, such a process is equivalent to a dephasing error on the measured photon, leading to a further loss of coherence due to the loss of off-diagonal terms in the state of the photon ($T_2$~noise).

In conclusion, if one encodes each node of the bucket brigade qRAM in an error correcting code, then all nodes of the circuit are activated at a physical level, and essentially the qRAM architecture becomes equivalent to a fanout architecture.
Even it the latter case, designing a good quantum error correcting code is highly non-trivial. An important issue is that the syndrome measurement should not reveal any information whatsoever about the physical location of the nodes affected by errors. Otherwise, path information is being revealed, which decoheres the system.

\section{Conclusions and open questions\label{sct:Conclusions}} 
We analyzed the robustness of the bucket brigade qRAM scheme introduced in \cite{PhysRevLett.100.160501, PhysRevA.78.052310} under an optimistic error model. 

\delete{The primary advantage of the bucket brigade scheme is the need for a polynomial in $n$ (rather than exponential) number of gate activations per memory reading. 
Yet, we give evidence for the hypothesis that for realistic error models, whenever the qRAM is used as a oracle for quantum searching, its error rate per gate has to scale as $o(2^{-n/2})$. Such an error rate is exponentially smaller than the error $\OC(1/n^2)$ proposed in \cite{PhysRevA.78.052310} (which is sufficient for algorithms with low query complexity), motivating the need for quantum error correction.}

\add{The primary advantage of the bucket brigade scheme is the need for a polynomial in $n$ (rather than exponential) number of gate activations per memory reading. When used for quantum algorithms \cite{PhysRevLett.103.150502, PhysRevLett.109.050505, quantph.1307.0411, PhysRevLett.113.130503, Nature.10.631, quantph.1408.3106}
 that require only a polynomial number of queries, the error rate of the qRAM can scale polynomially in $n$, and error correction may not be required. By contrast, we give evidence that under realistic error models, whenever the qRAM is used as an oracle for quantum searching, its error rate per gate has to scale as $o(2^{-n/2})$ \cite{PhysRevA.78.052310}, motivating the need for quantum error correction.}

We argued that using traditional error correcting techniques offsets the main advantage of the bucket brigade scheme when used with algorithms that make super-polynomially many oracle queries. Since each component of the routing architecture has to be actively error corrected in order to protect against detrimental faults, the overall scheme requires an exponential number of physical gate activations, even if the number of logical gate activations remains polynomial. 

An interesting open question is the existence of a realistic architecture-specific error correction technique that could recover the polynomial number of physical gate activations of the routing scheme while still guaranteeing fault-tolerance. For example, if one tries to use an error correction mechanism whereby one only uses multi-qubit code states along the active path, then one has the problem of extracting syndromes and applying corrections in a way that does not identify which path has the non-trivial syndromes (since such information would lead to decoherence). If in this case, for example, one attempts to extract the syndrome without leaving a trace of which node in a given level it came from, then the problem seems at least as challenging as implementing a reliable qRAM.

Moreover, it would be interesting to investigate whether the requirement for a super-polynomial suppression of the error rate is a characteristic of quantum searching algorithms or a more general feature of query complexity with faulty oracles.

\begin{acknowledgments}   
We thank Daniel Gottesman, Stacey Jeffery, Seth Lloyd, Lorenzo Maccone and Matteo Mariantoni for numerous insightful discussions and suggestions. We thank Seth Lloyd for pointing out to us nice examples of important algorithms that only use polynomially many queries. \add{We also thank the anonymous referees for their suggestions in improving the manuscript.} The authors were supported in part by NSERC, CIFAR, FQRNT, OGS, CryptoWorks21 and Industry Canada. IQC and PI are supported in part by the Government of Canada and the Province of Ontario. S.A. is grateful to Ronald de Wolf's  ERC Consolidator Grant QPROGRESS and Michele Mosca for support from NSERC.
\end{acknowledgments}

\appendix

\section{A simple decoherence model}\label{apdxA}
Let  us consider the error model considered in \cite{Regev:2008:IQS:1427895.1427975},
\begin{equation}\label{Aeqn1}
\RC_{ p}(\rho):=(1- p)\hat O \rho \hat O^\dagger +  p \rho,
\end{equation}
with $\hat O$ denoting the perfect oracle for quantum searching  and let us define
\begin{equation}\label{Aeqn2}
\DC_{ q}(\rho):=(1- q)\rho  +  q \vec{X}\rho \vec{X}^\dagger
\end{equation}
as the multi-qubit bit-flip channel where $\vec{X}$ is a shorthand notation for a tensor product of $\sigma_X$ bit-flip operators acting on some fixed subset of the qubits. The proof technique presented below for $\DC_q$ also applies to the case of multi-qubit dephasing channels).

The error model proposed in this paper (see Eq.~\eqref{eqn:9}) is 
\begin{equation}\label{Aeqn3}
O(\rho):=p_{rp}\hat O\rho \hat O^\dagger + p_{wp}\EC_{wp}(\rho) + p_{np}\EC_{np}(\rho).
\end{equation}
We show that the composition $\RC_p\circ \DC_q$ resembles (although not exactly the same) our error model Eq.~\eqref{eqn:9}, for suitable chosen $p$ and $q$. It follows immediately that the $\Omega(N)$ lower bound for the searching algorithm considered in  \cite{Regev:2008:IQS:1427895.1427975} is also a lower bound for the composition $\RC_p\circ \DC_q$, since channel composition cannot decrease the query complexity (one can simply incorporate $\DC_q$ into an appropriate unitary for the $\RC_p$ algorithm). A simple calculation yields:
\begin{widetext}
\begin{align}\label{Aeqn4}
\RC_p\circ\DC_q(\rho)&=(1- p)\hat O\DC_q(\rho)\hat O^\dagger  +  p \DC_q(\rho)\notag\\
&= (1-p)(1-q)\hat O\rho\hat O^\dagger + (1-p)q\hat O (\vec{X} \rho \vec{X}^\dagger) \hat O^\dagger  + p(1-q)\rho + p q \vec{X}\rho \vec{X}^\dagger  \notag\\
&=(1-p)(1-q)\hat O\rho\hat O^\dagger + (1-p)q\hat O (\vec{X} \rho \vec{X}^\dagger)\hat O^\dagger + p \DC_q(\rho).
\end{align}
\end{widetext}
We now identify the coefficients in Eq.~\eqref{Aeqn3} and Eq.~\eqref{Aeqn4}
\begin{empheq}[left=\empheqlbrace]{align}
p_{rp} &= (1-p)(1-q) \notag\\
p_{wp} &= (1-p)q \notag\\
p_{np} &= p,\label{Aeqn5}
\end{empheq}
and note that for any given probabilities $p_{rp}, p_{wp}, p_{np}$ satisfying $p_{rp} + p_{wp} + p_{np} = 1$, equations Eq.~\eqref{Aeqn5} have the solution
\begin{empheq}[left=\empheqlbrace]{align}
p&=p_{np}\notag\\
q&=\frac{p_{wp}}{p_{wp}+p_{rp}}.\label{Aeqn6}
\end{empheq}
We can therefore write
\begin{equation}\label{Aeqn7}
\RC_p\circ\DC_q(\rho)=p_{rp}\hat O\rho\hat O^\dagger + p_{wp}\hat O (\vec{X} \rho \vec{X}^\dagger)\hat O^\dagger + p_{np}\DC_q(\rho).
\end{equation}

Comparing Eq.~\eqref{Aeqn3} and Eq.~\eqref{Aeqn7}, we observe that the term $\hat O (\vec{X} \rho \vec{X}^\dagger)\hat O^\dagger$ is very similar to our wrong path term $\EC_{wp}(\rho)$ (the error that corresponds to reading out an incorrect memory location). The last term term $\DC_q(\rho)$ in Eq.~\eqref{Aeqn7} is not of the form of our no-path error term $\EC_{np}(\rho)$, as the latter consists of depolarizing channels of different strengths depending on the position of the address qubit (i.e., the qubits are affected in decreasing order of significance, that is, the first qubit is affected the most, whilst the last one the least). However, $\DC_q(\rho)$ is a decohering term, which seems to be a ``weaker" form of noise than $\EC_{np}(\rho)$. We showed above that even with this weaker decoherence term the quadratic speedup of any searching algorithm is lost. Therefore we have strong reasons to believe that adding a stronger decoherence term will not lower the quantum query complexity for the quantum searching problem. A rigorous proof of this conjecture remains an open problem.

\section{Error correction schemes}\label{apdxB}
\subsection{Correcting simple bit-flip errors}\label{apdxB1}
We show below that for a qRAM governed by a toy error model of the form
\begin{equation}\label{Beqn1}
O (\rho) = (1-p) \hat O \rho \hat O^{\dagger} + p \hat O (\vec X \rho \vec X^\dagger) \hat O^{\dagger},
\end{equation}
 the query error rate can be made arbitrarily small by using quantum error correction. Here $\hat O$ denotes the perfect oracle and $\vec X$ represents a multi-qubit bit-flip channel (a tensor product of individual bit-flip operators acting on an arbitrary subset of qubits). While such error models are not realistic for the architecture presented in this work, it may be that future designs allow for simpler error propagation. Such schemes could benefit from quantum error correction to sufficiently reduce their error rate to enable Grover search.

As Grover's algorithm requires $\OC(\sqrt{N})$ steps, one desires a target logical error rate of $\delta=\OC(1/\sqrt{N})$.
Since the faulty oracle has an error model that consists of a bit flip channel followed by the perfect oracle call, one can use a quantum error correcting code and apply the oracle in parallel along the qubits composing the code. The parallelism of the oracle calls mimics majority counting and allows for error correction to be performed between logical oracle call steps. For simplicity, we provide an example that corrects against bit flip errors only using the repetition code, however such an analysis could be extended to correct for phase flips using code families such as the color codes~\cite{PhysRevLett.97.180501}, higher dimensional Shor codes~\cite{PhysRevA.52.R2493}, or triorthogonal codes~\cite{PhysRevA.86.052329}. 

For example, consider an oracle of the form~$\ket{a} \ket{b} \rightarrow \ket{a} \ket{b \oplus f (a)}$, where $a, \ b \in \{0,1\}$. A logical oracle call that uses an $n$-qubit repetition code behaves as follows for states in the computational basis:
\begin{align}
\ket{a}\ket{b} \stackrel{V}{\rightarrow} \ket{a}^{\otimes n} \ket{b}^{\otimes n} \stackrel{\hat O^{\otimes n}}{\longrightarrow} \ket{a}^{\otimes n} \ket{b \oplus f(a)}^{\otimes n},
\end{align}
where $V$ denotes the isometric encoding. Therefore, given a repetition code of length~$d$, the code corrects for all errors up to $d/2-1$ physical bit flips by majority counting, using non-destructive $Z$-type stabilizer measurements. Therefore, the logical error rate becomes $p_L = p^{d/2}$. Choosing $d$~large enough allows the logical error rate to satisfy $p_L=p^{d/2}<\delta$, where $\delta$ is the desired target fidelity. Therefore
\begin{equation}\label{Beqn3}
d >  \dfrac{2\log{\delta}}{\log p} = \dfrac{ 2 \log{\left(1/\sqrt{N}\right)}}{\log{p}} = \dfrac{ n}{\log{\left(1/p\right)}}.
\end{equation}
Each of the $n$~address qubits that serve as input to the oracle call must be encoded into a repetition code of length~$d$. Hence, the total number of oracle calls for the complete Grover search algorithm is~$\OC(nd\sqrt{N})  = \mathcal{O} (n^2 \sqrt{N}) = \mathcal{O} (\sqrt{N} (\log{N})^2 )$. As such, there is a logarithmic penalty for error correction, yet the scaling is not linear as in the error model of Regev and Schiff~\cite{Regev:2008:IQS:1427895.1427975}.  

\subsection{The failure of repetition codes for Regev and Schiff error model}\label{apdxB2}
The above error correction scheme is not applicable to the error model presented in~\cite{Regev:2008:IQS:1427895.1427975}, described by~$\mathcal{R}_p (\rho) = (1-p) \hat O \rho \hat O^{\dagger} + p \rho$, since the failure of an oracle call can lead to an uncorrectable error, as demonstrated below. Consider the following example of the 3-qubit repetition code, where rather than all three oracles calls succeeding, the oracle call on the first set of qubits fails. The computational states evolve as:
\begin{align}
\ket{000}\ket{000} &\stackrel{\hat O_2 \hat O_3}{\longrightarrow} \ket{000}\ket{0 f(0)f(0)}\label{Beqn4}\\
\ket{111}\ket{000} &\stackrel{\hat O_2 \hat O_3}{\longrightarrow} \ket{111}\ket{0 f(1)f(1)}.\label{Beqn5}
\end{align}
Consider the action of such a faulty oracle on the encoded state $(\ket{000}+\ket{111})/\sqrt{2}$, for $f(0) = 0$ and~$f(1) = 1$ . The 
resulting mapping is
\begin{align}
&\dfrac{1}{\sqrt{2}} (\ket{000}+\ket{111})\otimes \ket{000}\notag\\
&\qquad \stackrel{\hat O_2 \hat O_3}{\longrightarrow} \dfrac{1}{\sqrt{2}} (\ket{000}\ket{000}+\ket{111}\ket{011}).
\end{align}
The syndrome check operators for the repetition code are the parity check operators $\{Z_1 Z_2, Z_2 Z_3\}$. They are used to determine if an oracle call has failed by measuring the ancilla qubits. However, the measurement collapses the state to either~$\ket{000}\ket{000}$ or~$\ket{111}
\ket{011}$. 
Upon applying the appropriate correction based on the measured syndromes, the resulting state becomes either $\ket{000}\ket{000}$ or $\ket{111}\ket{111}$. Therefore, the logical oracle call has failed, since the correct result must yield the superposition~$(\ket{000}\ket{000}+\ket{111}\ket{111})/\sqrt{2}$.

As expected, the error correction properties of the repetition code are not in violation of the results of Ref.~\cite{Regev:2008:IQS:1427895.1427975}, which state that a linear number of noisy black-box oracle calls are required, even with the addition of error correction. 

\subsection{The failure of repetition codes for our error model}\label{apdxB3}
Consider the oracle error model:
\begin{equation}
O(\rho):=p_{rp}\hat O\rho \hat O^\dagger + p_{wp}\EC_{wp}(\rho) + p_{np}\EC_{np}(\rho),
\end{equation}
where $\hat O$ is the perfect oracle call while $\EC_{wp}(\rho)$ and $\EC_{np}(\rho)$ are the \emph{wrong path} and \emph{no-path} terms, respectively. We model the \emph{wrong path} term as a convex combinations of bit-flip channels followed by perfect oracle calls. An example of one of those terms is the second term in Equation~\ref{Beqn1}. We model the \emph{no-path} term as taking any input state and mapping it to a fixed state~$\ket{g}$, which represents the loss of a qubit to be replaced by any fixed ancillary state. It should be noted that in the \emph{no-path} case, the readout ancilla state does not change. Consider the action of the noisy channel on the five-qubit repetition code. Each instance of the channel has a certain probability of failure given by the associated weights. Focusing on one particular instance where the first address photon is lost and the second is affected by a bit flip, the resulting mapping on the computational basis states is given by:

\begin{align}
\ket{00000}\ket{00000} &{\longrightarrow}
\ket{g1000}\ket{0 f(1)f(0)f(0)f(0)}\label{Beqn8}\\
\ket{11111}\ket{00000}  &{\longrightarrow}
\ket{g0111}\ket{0 f(0)f(1)f(1)f(1)}.\label{Beqn9}
\end{align}

Again choosing $f(0) = 0$ and $f(1) = 1$, a superposition of input states in the computational basis evolves as
\begin{align}
&\dfrac{1}{2} \mathrm{P}\left[ \left( \ket{00000}+\ket{11111}\right) \otimes \ket{00000}\right] {\longrightarrow}\notag\\
&\qquad \dfrac{1}{2}\left(\mathrm{P}\left[\ket{g1000}\otimes\ket{01000}\right]+\mathrm{P}\left[\ket{g0111}\otimes\ket{00111}\right]\right)\label{Beqn10},
\end{align}
where $\text{P}[\bullet]$ denotes the projector onto its argument.
The measurement of the stabilizers of the 5-qubit code on the ancillary states results in the collapse of the state into one of two terms depending on the syndrome measured. Note that the \emph{no-path} term is the term that destroys coherence, similarly to the error term in the Regev and Schiff model~\cite{Regev:2008:IQS:1427895.1427975}.

%\bibliographystyle{apsrev4-1} % no need for this if longbibliography class is used
%\bibliography{./qram}


\end{document}